\documentclass[12pt]{article}

\usepackage{jheppub} 

\usepackage{amsmath,amsfonts}
\usepackage{amssymb}
\usepackage{mymacros}
\usepackage{bbm}
\usepackage{braket}
\usepackage{slashed} 

\usepackage{tikz}
\usetikzlibrary{decorations.pathmorphing, decorations.markings, arrows.meta}

\usepackage{tensor}

\renewcommand{\sl}{\slashed}

\newcommand{\ms}{\mathbb} 

\newcommand{\R}{\ms R}
\renewcommand{\T}{\ms T}

\newcommand{\Q}{\ms Q}
\newcommand{\U}{\ms U}
\renewcommand{\B}{\ms B}

\newcommand{\m}{\mu}
\newcommand{\n}{\nu}
\renewcommand{\t}{\tau}
\renewcommand{\r}{\rho}
\renewcommand{\l}{\lambda}
\renewcommand{\a}{\alpha}
\renewcommand{\b}{\beta}
\renewcommand{\d}{\delta}
\renewcommand{\g}{\gamma}
\renewcommand{\k}{\kappa}

\newcommand{\Kfree}{K_{\rm free}}

\newcommand\tra{^{\mathpalette\raiseT\intercal}}

\newcommand\raiseT[2]{%
\setbox0\hbox{$#1{#2}$}\raise\dp0\box0}

\title{
Heat Kernel Methods for Multiloop Calculations in Curved Spacetime: Nonzero Spin }
\author[1,2]{Igor Carneiro}
\author[1]{Gero von Gersdorff}
\affiliation[1]{Department of Physics, Pontifícia Universidade Católica do Rio de Janeiro,
Rua Marquês de São Vicente 225, 22451-900, Rio de Janeiro, Brazil}
\affiliation[2]{École Polytechnique, Institut Polytechnique de Paris,
Rte de Saclay, 91120 Palaiseau, France}

\emailAdd{gersdorff@puc-rio.br}
\emailAdd{igor.carneiro-dos-santos@polytechnique.edu}

\abstract{In a previous paper we have presented a general formalism for computing Feynman diagrams for scalar fields in curved spacetime at any loop order using heat kernel methods. The main technique used is the expansion of the fully off-diagonal heat kernel in Riemann normal coordinates. In this work we extend this to include fields of nonzero spin, in particular spin-$\frac{1}{2}$ fermions. }

\begin{document}

\maketitle

\section{Introduction}

The heat kernel (HK) is an important tool in quantum field theory that has found numerous applications, including the calculation of the effective action, renormalization group equations of local operators, effective field theory of heavy particles, anomalies, chiral Lagrangians, to name a few. Moreover, it is  indispensable for  computing the behavior of a quantum field theory in classical gauge or gravitational backgrounds.
One of the  advantages of the method is that it is formulated in a manifestly gauge and diffeomorphism invariant way. 

Most of the applications of the method in the above contexts have focused on the one-loop case \cite{Fock:1937dy,schwinger1951,dewitt1965dynamical,DeWitt:1967ub,Gilkey:1975iq,Barvinsky:1985an,Avramidi:1990ug,Avramidi:1990je,Fujikawa:1979ay,Fujikawa:1980eg,Gasser:1983yg,Ball:1988xg,vonGersdorff:2003dt,vonGersdorff:2006nt,Hoover:2005uf,Barvinsky:2005qi,vonGersdorff:2008df}
with few exceptions in the flat \cite{Duff:1975ue,Batalin:1976uv,Batalin:1978gt,Bornsen:2002hh,Ivanov:2020yrc,Ivanov:2022aco,Ivanov:2024lbs} and curved \cite{Bunch:1979uk,Luscher:1982wf,Kodaira:1985pg,vandeVen:1991gw,Bilal:2013iva,Carneiro:2024slt} backgrounds. 
However, until recently, a systematic formalism that enables one to write the local expansion of an arbitrary multi-loop Feynman graph was still lacking. 
In a series of papers we developed such a formalism \cite{vonGersdorff:2022kwj,vonGersdorff:2023lle,Carneiro:2024slt}. 
References \cite{vonGersdorff:2022kwj,vonGersdorff:2023lle} dealt with the flat case (i.e.,no gravitational backgrounds), including scalar, spin 1/2, and spin-one fluctuating fields. On the other hand, reference \cite{Carneiro:2024slt} dealt with arbitrary curved background but focused on spin-zero fluctuating fields for simplicity.
Thus, the case of arbitrary spin and arbitrary (curved) backgrounds is still missing. 
The current work  fills this gap.

Let us say we would like to compute the effective Lagrangian $\mathcal L_{\rm eff}(x)$ in an expansion in local operators in $ x$. To this end it is very convenient to adopt Riemann normal coordinates (RNCs) centered around the point $x$. We will generally denote such coordinates by $y$, and place the base point at the origin. Then, in the presence of background gauge fields $A_\mu$ (which includes the composite spin-connection field) it is  natural to adopt the radial gauge $y^\mu A_\mu=0$. All background quantities such as the metric, vielbein, gauge field, spin-connection etc then possess a Taylor expansion around the origin whose coefficients are tensors at the base point. 

The HK is essentially a two-point function, and thus  depends on two spacetime points $y$ and $y'$.
While RNCs have been used previously (see e.g.~\cite{Bunch:1979uk,Luscher:1982wf,Buchbinder:1983nug,Buchbinder:1984zba,Inagaki:1997kz,Toms:2014tia,Larue:2023uyv}), to the best of our knowledge all these expansions identified one of the end points as the base point.
However, in all but the simplest cases, this is insufficient, and instead one needs to expand in both arguments. This is one of the technical novelties of \cite{Carneiro:2024slt} and this work.

This paper is organized as follows. In section \ref{sec:props} we present the HK representations of fields of nonzero spin, in particular spin 1/2 and spin one, allowing for arbitrary gauge and gravitational backgrounds. 
In section \ref{sec:formalism} we lay out our formalism, which is a rather straightforward generalization of the flat, any spin \cite{vonGersdorff:2023lle} and curved, spin-zero \cite{Carneiro:2024slt} cases.
Section \ref{sec:HK} contains our main technical result, the expansion of the general HK in RNCs and radial gauge. We illustrate our formalism with a rudimentary three-loop example in section \ref{sec:example}, and in section \ref{sec:conclusions} we present our conclusions. In appendix \ref{sec:conventions} we summarize out notation. 
Appendices \ref{sec:momentumintegral} and \ref{sec:grav} contain summaries of previous results relevant for this work.

\section{Heat kernel representations of propagators of nonzero spin}
\label{sec:props}

\subsection{The heat kernel for general spin- and gauge-connections}

As there are no spinorial representations for diffeomorphisms, in order to construct a theory with fields of non-integer spin we must introduce a local vielbein 
\be
e\indices{_\mu^a}(x)e\indices{_\nu^b}(x)\eta\indices{_{ab}}\equiv g\indices{_{\mu\nu}}(x)\,.
\label{eq:vielbeindef}
\ee 
The vielbein is only defined up to local Lorentz transformations, $e\indices{_\mu^a}(x)\to 
\Lambda\indices{^a_b}(x)e\indices{_\mu^b}(x)$. Fermions can then be introduced as standard spinorial representations of the Lorentz group.
In this and the following sections, we denote curved indices by lowercase Greek letters and local Lorentz  indices by lowercase Latin letters. 
The two can be converted via the vielbein $e\indices{_\mu^a}$ and its inverse $e\indices{_a^\mu}$, for instance $\gamma_\mu^{}=e\indices{_\mu^a}\gamma\indices{_a}$ where $\gamma_a$ are the standard Dirac matrices satisfying $\{\gamma_a,\gamma_b\}=\eta_{ab}$.\footnote{We use particle physicists' $+---$ convention for the curved and flat metrics $g_{\mu\nu}$ and $\eta_{ab}$.} 
We define the standard spin connection by
\be
\omega\indices{_\mu^a_b}
\equiv -e\indices{_b^\nu}\nabla\indices{_\mu} e\indices{_\nu^a}\,,
\label{eq:spincon}
\ee
where in our notation $\nabla $ is covariant with respect to curved indices only.

The HK is defined by 
\be
K(t,X,x,x')\equiv e^{-it(D^2+X+m^2)}\delta(x,x')\,,
\ee
where $\delta(x,x')\equiv [-\det g(x)]^{-1/2}\delta(x-x')$ is the biscalar delta function and the covariant derivative $D_\mu$ contains Christoffel, gauge, and spin connections
\be
D_\mu=\nabla_\mu+\mathcal A_\mu+\omega_\mu\,.
\ee
Here,  $\mathcal A_\mu$   is the matrix-valued, anti-Hermitian gauge connection,
and 
\be
\omega_\mu=\frac{1}{2}\omega_{\mu ab}\Sigma_{s}^{ab}\,,
\ee
is the matrix valued spin connection with the spin-$s$ generators $\Sigma_s^{ab}$.

We now switch to RNCs, which we will denote by $y$ and whose base point coincides with the origin $y=0$.
We can then split the HK into a free HK and a background-field dependent piece as in \cite{Carneiro:2024slt}:
\be
K(t,X,y,y')\equiv \Kfree(t,y,y')B(it,X,y,y')\,,
\label{eq:splitting}
\ee
where
\be
 \Kfree(t,y,y')\equiv i(4\pi i t)^{-\frac{d}{2}}e^{-i\frac{(y-y')^2}{4t}}=
 \int_k e^{itk^2- ik(y-y')}\,,
\ee
is the free, massless HK.
Notice that $\Kfree$ and hence the decomposition in eq.~(\ref{eq:splitting}) are coordinate dependent  and are thus only valid in RNCs.\footnote{As different RNCs with the same base point are related by a rigid Lorentz transformation, $\Kfree$ takes the same form in all such frames. For the time being we do not need any more properties of  RNCs, they will be relevant only when  performing the actual computation of $B(\t,X,y,y')$ in section \ref{sec:HK}.}
Here and in the following we use the shorthands
\be
\int_{x}\equiv \int  d^dx\,,
\qquad
\int_{k}\equiv \int  \frac{d^dk}{(2\pi)^d}\,,
\qquad
\int_{t}\equiv \int_0^\infty  dt\,.
\ee

The function $B(\t,X,y,y')$, {\em defined} by  eq.~(\ref{eq:splitting}), is calculated in section \ref{sec:HK} in an expansion about $\t=0$, $y=0$, and $y'=0$. For the remainder of this section we assume it is known and proceed to express the propagators of various spins in terms of it.

\subsection{Spin zero}

For completeness we review here the spin-0 case also. Its propagator has the HK representation
\be
\frac{-i}{D^2+m^2}\delta(x,x')=\int_t  K(t,0,x,x')\, e^{-itm^2}\,.
\ee
In RNCs, the propagator thus becomes 
\be
\frac{-i}{D^2+m^2}\delta(y,y')=\int_{k,t} e^{itk^2 - ik(y-y')}e^{-itm^2}  B(it,0,y,y')\,.
\label{eq:derivativeprop}
\ee
Propagators of derivatives of scalar fields, such as $\braket {D_\mu\phi(x) \phi^\dagger(x')}$ can be written as
\be
D_\mu\frac{-i}{D^2+m^2}\delta(y,y')=\int_{k,t} e^{itk^2 - ik(y-y')}e^{-itm^2}  ( D_\mu-ik_\mu)B(it,0,y,y')\,.
\ee

\subsection{Spin 1/2}

The background-field dependent propagator for a spin-$\frac12$ field can be parametrized as
\be
\frac{i}{i\sl D-m}\delta(x,x')=(i\sl D+m)\frac{-i}{D^2+X_{\frac{1}{2}}+m^2}\delta(x,x')\,,
\ee
where $\sl D=\gamma^\mu D_\mu$.
Using the HK in RNCs, see eq.~(\ref{eq:splitting}), we therefore write
\be
\frac{i}{i\sl D-m}\delta(y,y')
=\int_{k,t} e^{itk^2 - ik(y-y')}e^{-itm^2}  (i\sl D+\sl k+m)B(it,X_\frac{1}{2},y,y')\,,
\label{eq:propferm}
\ee
where the background-field dependent mass matrix $X_\frac{1}{2}$ reads
\be
X_\frac{1}{2}\equiv \frac{1}{2}\gamma^a\gamma^b[D_a,D_b]\equiv \frac{1}{2}\gamma^a\gamma^bF_{ab}\,,
\ee
with 
\be
F_{ab}=\mathcal F_{ab}+\omega_{ab}=\mathcal F^{i }_{ab}{\mathfrak t_i}+\frac{1}{2}R_{ab cd}\Sigma_\frac{1}{2}^{cd}\,,
\ee
where the ${\mathfrak t_i}$ are the anti-Hermitian generators of the gauge group (equal to $i\times$charge for Abelian groups), and the spin-$\frac{1}{2}$ generators are defined as
\be
\Sigma_\frac{1}{2}^{ab}\equiv \frac{1}{4}\left[\gamma^a,\gamma^b\right]\,.
\ee
Using that $\gamma^a\gamma^b\gamma^c\gamma^d R_{abcd}=-2 R$ by the symmetries of the Riemann tensor this simplifies to
\be
X_\frac{1}{2}=\frac{1}{2}\gamma^a\gamma^b\mathcal F_{ab}-\frac{1}{4}R\,.
\label{eq:X12}
\ee

\subsection{Spin one}

Next, we move to  the representation of a massless spin-1 field  in the covariant Feynman gauge,
\footnote{The gauge fixing term reads $\mathcal L_{\rm gf}=-\frac{1}{2}(D_\mu \mathcal A_{i}^\mu)^2$ where $D_\mu$ only contains background fields.} whose propagator reads
\be
\frac{i}{D^2+X_1}\delta(y,y')
=-\int_{k,t} e^{itk^2 - ik(y-y')}e^{-itm^2}  B(it,X_1,y,y')\,,
\ee
where $X_1= \Sigma_{1}^{ab}(\mathcal F_{ab}+\frac{1}{2}\omega_{ab})$ with 
$(\Sigma_1^{ab})_{cd}=\delta^a_c\delta^b_d-\delta^a_d\delta^b_c$. Explicitly
\be
(X_1)^{ai}_{\ bj}
=2\eta^{ac}\mathcal F^{k}_{cb}f_{kj}^{\ \ i}-R^a_{\ b}\delta^i_{\ j}.
\label{eq:X1}
\ee
The associated ghost propagators is given by that of a complex adjoint scalar with $X_0=0$.
Propagators of derivative of vector fields, which appear in non-Abelian gauge theories, can be constructed as in the scalar case, eq.~(\ref{eq:derivativeprop}).

One can add to $X_1$ further terms  arising from additional backgrounds (e.g., scalar fields). 
Such an inclusion is straightforward \cite{vonGersdorff:2022kwj} and will not be shown explicitly here.

\section{The master formula for arbitrary Feynman graphs } 
\label{sec:formalism}

In this section we generalize previous results, the flat space formalism for arbitrary spin fields \cite{vonGersdorff:2022kwj,vonGersdorff:2023lle} and the curved space formalism for spin zero fields \cite{Carneiro:2024slt} to find the most general master formula valid for arbitrary spin in curved space. 

We define the field-dependent couplings
\be
C(x)\equiv \prod_\Psi \frac{\partial}{\partial \Psi} \sqrt{g} \mathcal L_{\rm int}\,,
\ee
where the product goes over all fluctuation fields contributing to the vertex. We include the possibility that some of these fields contain derivatives, 
e.g.~$\Psi=D_\mu\Phi$ etc., and $\mathcal L_{\rm int}$ is the (scalar) Lagrangian containing the interaction in question.

The effective action (up to symmetry factors and signs for closed fermion loops) is given by
\be
S_{\rm eff}=i^{V-1}\int_{\{x_n\}} \prod _{i=1}^P G_i\prod_{n=1}^V C_n\,,
\ee
where the $G_i$ are the propagators of the $i^\text{th} $ edge in position space, and the $C_n$ are the  couplings associated to the $n^\text{th}$  vertex. Both $C_n$ and $G_i$ are background-field dependent. 
All gauge and spin indices are suppressed here, it is understood that they are contracted as dictated by the diagram (the final result has no free indices).
For the time being, we consider a general frame with coordinates denoted by $x$.

The effective Lagrangian, defined as $S_{\rm eff}=\int_x \sqrt {g(x)} \, \mathcal L_{\rm eff}(x)$ can be written in terms of an arbitrary  function $\tilde x$ of the $V$ vertex coordinates $x_n$ as
\be
\mathcal L_{\rm eff}(x)=i^{V-1}\int_{\{x_n\}}\delta(x,\tilde x(x_n))\prod_n C_n\prod _i G_i\,.
\ee
We will now chose $\tilde x(x_n)=x_1$, and also make a change of variables to RNCs $y_n$ with base point $x$ (conveniently taken to be at the origin of the RNCs) to get
\be
\mathcal L_{\rm eff}=i^{V-1}\int_{\{y_n\}}\delta(y_1)\prod_n C_n\prod _i G_i\,.
\ee

Next, we implement the parametrization of the propagators in terms of the HK of section \ref{sec:props}.
From there it is clear that all propagators possess a representation
\be
G(y,y')=
\int_{k,t} e^{itk^2 - ik(y-y')}e^{-itm^2}  E(it,k,y,y')\,,
\ee
in particular, $E(\t,k,y,y')=B(\t,X_0,y,y')$ for scalars, $E(\t,k,y,y')=(i\sl D+\sl k+m)B(\t,X_\frac{1}{2},y,y')$ for spin-1/2 fermions, and
$E(\t,k,y,y')=-B(\t,X_1,y,y')$ for vectors in Feynman gauge. 
Notice that for the case of spin-1/2 fermions, $E$  depends not only on the vertex positions $y$, $y'$ but also on the propagator momentum $k$, and the same is true for propagators of derivatives of fields, see e.g.~eq.~(\ref{eq:derivativeprop}). This, together with the dependence of $B$ on the spin and gauge connection, is the main difference to the purely scalar case of ref.~\cite{Carneiro:2024slt}. 
We now follow refs.~\cite{vonGersdorff:2022kwj,vonGersdorff:2023lle} in order to deal with this $k$- dependency in a completely universal way.
We  define the quantity
\be
\Gamma(\t_i,y_n,k_i)\equiv \prod_n C_n\prod_i E_i\, e^{-\t_i m_i^2}\,.
\label{eq:Gammadef}
\ee
Note that $\Gamma$ is  invariant wrt.~internal and Lorentz gauge transformations but  not diffeomorphism invariant (due to the use of RNCs). However, the coefficients of its expansion about the base point do transform as tensors.
The diagram in question now contributes to the effective Lagrangian as
\be
\mathcal L_{\rm eff}=(-i)^L \int_{\{\t_i,k_i, y_n\}}\delta(y_1)e^{-iy_n\B_{ni}k_i+\t_ik_i^2}\,\Gamma(\t_i,y_n,k_i)\,,
\ee
where $\B$ is the directed incidence matrix of the graph (see eq.~(\ref{eq:Bni})).
The integration over the vertex positions $y_n$ and propagator momenta $k_i$ can then be performed analytically as follows.
The function
\be
I'(\t_i,y_n,k_i)\equiv(-i)^L\delta(y_1)e^{-iy_n\B_{ni}k_i+\t_ik_i^2}\,,
\ee
has a Fourier transform $ I'(\t_i,p_n,z_i)$ (with respect to both $y_n$ and $k_i$) that has been calculated in ref.~\cite{vonGersdorff:2023lle} in terms of graph polynomials and which we review in appendix  \ref{sec:momentumintegral}.
In terms of $ I'(\t_i,p_n,z_i)$ we can write a very simple master formula
\be
\mathcal L_{\rm eff}= \int_{\{\t_i\}}I'(\t_i,i\tfrac{\partial}{\partial y_n},-i\tfrac{\partial}{\partial k_i})\,\Gamma(\t_i,y_n,k_i)\bigr|_{y_n=0,k_i=0}\,.
\label{eq:master}
\ee
Notice that only the integrals over the Schwinger parameters $\t_i$ remain. In particular, the usual momentum integration has been done in  the form of the  Fourier transform. The function $\Gamma$ is polynomial in the $k_i$ and can be systematically expanded around $y_n=0$ and $\t_i=0$, with coefficient being gauge and diffeomorphism invariant operators  at the origin.
The main remaining technical difficulty is the expansion of the function $B(\t,X,y,y')$ around $y=y'=0$, and $\t=0$, which we turn to in the next section.

\section{The expansion of the heat kernel in Riemann normal coordinates and radial gauge}
\label{sec:HK}


In order to simplify calculations and conveniently implement a gauge fixing, we introduce RNCs, in terms of which the metric reads
\begin{align}
	g_{\mu\nu}(y)={}&\eta_{\mu\nu}-\frac{1}{3}R_{\mu\alpha\nu\beta}y^{\alpha}y^{\beta}-\frac{1}{6}R_{\mu\alpha\nu\beta;\gamma}y^{\alpha}y^{\beta}y^{\gamma}
\nn\\
{}&
+\left(\frac{16}{15}\eta^{\r\s}R_{\m\a\r\b}R_{\n\g\s\d}
		-\frac{6}{5}R_{\m\a\n\b;\g\d}    
		 \right)\frac{1}{24}y^\a y^\b y^\g y^\d +\dots
\label{eq:metricRNC}
\end{align}
As before, we denote points in RNCs by the letter $y$ to distinguish them from general frames for which we use the letter $x$. The curvature tensors on the right hand side are evaluated at the origin $y=0$.
Similar expansions can be found for all kinds of background fields, see for instance refs.~\cite{willmore1993riemannian,Brewin:2009se,Carneiro:2024slt}.

We would like to find similar expansion for the vielbein. Due to the local Lorentz gauge freedom, 
eq.~(\ref{eq:vielbeindef}) does however not completely fix this expansion.
Local Lorentz transformations are similar in nature to ordinary Yang-Mills transformations, with the important distinction that the connection $\omega_\mu$ defined in eq.~(\ref{eq:spincon}) is not an independent degree of freedom. 
For regular gauge connections, it is natural to combine RNCs with the so-called radial gauge, which sets to zero the spacetime radial projection of the connection
\begin{equation}
	y^{\mu}\mathcal A_\mu(y)=0\,.
\end{equation}
This leads to an expansion 
\begin{equation}
	\mathcal A_{\mu}(y)=-\frac{1}{2}\mathcal F_{\mu\alpha}y^{\alpha}-\frac{1}{3}\mathcal F_{\mu\alpha;\beta}y^{\alpha}y^{\beta}+\left(\frac{1}{24}R\indices{^\lambda_\alpha_\mu_\beta}\mathcal F\indices{_\lambda_\gamma}-\frac{1}{8} \mathcal F_{\mu\alpha;\beta\gamma}\right)y^{\alpha}y^{\beta}y^{\gamma}+\dots
	\label{eq:radialexp}
\end{equation}
We would like to extend this simple expansion to the spin connection. In fact, imposing
\be
y^\mu\omega_\mu(y)=0\,,
\label{eq:radialspin}
\ee
indeed fixes the gauge ambiguity in the definition of the vielbein, which is now deterimned up to global Lorentz transformations. The expansions read
\begin{align}
e\indices{_\mu^a}(y)={}&\delta\indices{_\mu^a}-\frac{1}{6}R\indices{^a_\alpha_\mu_\beta}y^{\alpha}y^{\beta}-\frac{1}{12}R\indices{^a_\alpha_\mu_{\beta;}_\gamma}y^{\alpha}y^{\beta}y^{\gamma}
\nn\\{}&
+\left(\frac{1}{120}\eta^{\rho\sigma}R\indices{^a_\alpha_\rho_\beta}R\indices{_\mu_\gamma_\sigma_\delta}
-\frac{1}{40}R\indices{^a_\alpha_\mu_{\beta;}_\gamma_\delta}\right)y^{\alpha}y^{\beta}y^{\gamma}y^{\delta}+\dots
\label{eq:eexp}
\\
\label{eq:einvexp}
e\indices{^\mu_a}(y)={}&\delta\indices{^\mu_a}+\frac{1}{6}R\indices{_a_\alpha^\mu_\beta}y^{\alpha}y^{\beta}+\frac{1}{12}R\indices{_a_\alpha^\mu_{\beta;}_\gamma}y^{\alpha}y^{\beta}y^{\gamma}
\nn\\{}&
+\left(\frac{7}{360}\eta^{\rho\sigma}R\indices{_a_\alpha_\rho_\beta}R\indices{^\mu_\gamma_\sigma_\delta}+\frac{1}{40}R\indices{_a_\alpha^\mu_{\beta;}_\gamma_\delta}\right)y^{\alpha}y^{\beta}y^{\gamma}y^{\delta}+\dots
\end{align}
It may be checked explicitly that the expansions satisfy both the RNC condition eq.~(\ref{eq:metricRNC}) via eq.~(\ref{eq:vielbeindef}) as well as the local radial gauge condition (\ref{eq:radialspin}) of the spin connection, eq.~(\ref{eq:spincon}).
As a result, we may now simply use the expansion eq.~(\ref{eq:radialexp}) for the spin connection directly and bypass the expansion of the vielbein.\footnote{Whereas eq.~(\ref{eq:radialexp}) is sufficient for the calculation of $B(\t,X,y,y')$, the expansion of the vielbein eqns.~(\ref{eq:eexp}) and eq.~(\ref{eq:einvexp}) are still needed in other places, see the explicit example in section \ref{sec:example}.}



In order to compute the function $B(\tau, X,y,y')$ as defined implicitly in eq.~(\ref{eq:splitting}),
we follow the same method as in ref.~\cite{Carneiro:2024slt}. 
We are after the expansion coefficients 
\be
\lim_{y,y'\to 0}\,\partial_{\mu_1}\dots\partial_{\mu_n} \partial'_{\nu_1}\dots
\partial'_{\nu_m} B(\tau, X,y,y')\,,
\label{eq:partials}
\ee
which in RNCs and radial gauge are tensors at the origin.
We take advantage of the fact that the quantities 
\be
\lim_{y'\to y}
D_{(\mu_1} \dots D_{\mu_n)} B(\tau, X, y,y')
\label{eq:covariant} 
\ee
 are known \cite{Decanini:2005gt,Groh:2011dw}, they are tensors at the point $y$. We then decovariantize these quantities (as well as its covariant derivatives acting outside the limit), take the limit $y\to 0$, and solve for the coefficients (\ref{eq:partials}).
 The  expansion coefficients (\ref{eq:partials}) are expressed in terms of the  quantities
 (\ref{eq:covariant}), as well as the connections and its derivatives, all evaluated at the origin. All of these are known, at least as an expansion in RNCs and radial gauge.
A very useful identity in the course of this evaluation is 
\be
B(\t,X,y,y')=B(\t,X,y',y)|_{A\to -A}\,,
\label{eq:flipxy}
\ee
where $A$ is the connection appearing in the covariant derivative, i.e., the sum of gauge and spin connections $A=\mathcal A_\mu+\omega_\mu$. This identity allows us to relate coefficients of $\mathcal O(y^ny'^m)$ to those of $\mathcal O(y^my'^n)$.

Even though the calculation outlined in the preceding paragraph is rather cumbersome in practice, there are no further conceptual difficulties.
When all the dust settles, the final result reads 
\begin{align}
\label{eq:Bfinal}
B(\tau,X,y,y')& =
B^{\rm grav}(\tau,X,y,y')
+\frac{\tau^2}{12}F_{\mu\nu}F^{\mu\nu}
-\frac{\tau}{6}F\indices{_{\mu\nu;}^\nu}(y^\mu-y'^\mu)
\nonumber\\
{}&
+\left(
\frac{\tau}{6}F\indices{_\lambda_{\mu;}^\lambda_{\nu}}
-\frac{\tau}{6}F\indices{_{\lambda\mu}}F\indices{^\lambda_{\nu}}
\right)\frac{1}{2}y^\mu y^\nu
+\left(
-\frac{\tau}{6}F\indices{_\lambda_{\mu;}^\lambda_{\nu}}
-\frac{\tau}{6}F\indices{_{\lambda\mu}}F\indices{^\lambda_{\nu}}\right)
\frac{1}{2}y'^\mu y'^\nu
\nonumber\\
{}&
+\left(
\frac{1}{2}F_{\mu\nu}
+\frac{\tau}{6}F\indices{_\lambda_{[\mu;}^\lambda_{\nu]}}
+\frac{\tau}{6}F\indices{_{\lambda(\mu}}F\indices{^\lambda_{\nu)}}
-\frac{\tau}{2}F_{\mu\nu}X
-\frac{\tau}{12}F_{\mu\nu}R
\right)y^\mu y'^\nu
\nonumber\\
{}&
+\left(
\frac{11}{12}F\indices{_{\lambda\mu}}R\indices{^\lambda_{\nu}_{\rho}_\sigma}
-\frac{1}{4}F_{\sigma\mu;\nu\rho}
-\frac{1}{12}F_{\sigma\mu}R_{\nu\rho}
\right)\frac{1}{6}(y^\mu y^\nu y^\rho y'^\sigma-y'^\mu y'^\nu y'^\rho y^\sigma)
\nonumber\\
{}&
+\left(-\frac{1}{2}F_{\rho \mu}F_{\nu\sigma}
-\frac{1}{3}F_{\mu \rho;\nu\sigma}
-\frac{1}{3}F_{\mu \rho;\sigma\nu}
\right.
\nn
\\
{}&
\left.
+\frac{1}{4}R\indices{^\lambda_{\mu}_{\nu}_{\rho}}F\indices{_{\sigma}_\lambda}
-\frac{1}{4}R\indices{^\lambda_{\rho}_{\sigma}_{\mu}}F\indices{_{\nu_\lambda}}
-\frac{1}{9}F_{\mu\rho}R_{\nu\sigma}
\right)\frac{1}{4}y^\mu y^\nu y'^\rho y'^\sigma
\nonumber\\
{}&
-\frac{1}{24\tau}F_{\alpha\beta} R_{\mu\rho\nu\sigma}\, y^\alpha y^\mu y^\nu y'^\beta y'^\rho y'^\sigma 
+\text{operators of dimension $>$ 4}.
\end{align}
Here, $F$ is  the sum of (matrix valued) gauge and spin connection field strengths, 
\be
F_{\mu\nu}\equiv [D_\mu,D_\nu]=\mathcal F_{\mu\nu}+\omega_{\mu\nu}, 
\ee
and $B^{\rm grav}$ contains all the purely gravitational terms already computed in ref.~\cite{Carneiro:2024slt}. For convenience, we reproduce this function in appendix \ref{sec:grav}. Note that these terms have some implicit dependence on the gauge and spin connections via eqns.~(\ref{eq:X12}) and (\ref{eq:X1}).

\section{A three-loop example}
\label{sec:example}

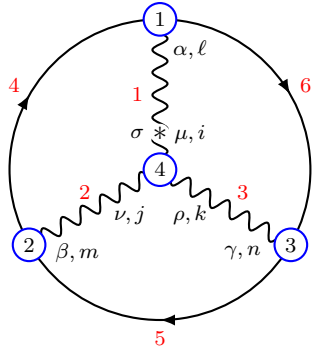
\begin{figure}
\begin{center}
\begin{tikzpicture}[thick, scale=2,
  midarrow/.style={
    postaction={decorate},
    decoration={markings,
      mark=at position 0.5 with {\arrow{Latex[length=2mm]}}}
  }]

  \draw (0,0) circle(1);

  \draw[decorate, decoration={snake, amplitude=1mm, segment length=3mm}] (0,0) -- (0,1);
  \node at (0.2,0.8) {\scriptsize$\alpha, \ell$};

  \draw[decorate, decoration={snake, amplitude=1mm, segment length=3mm}] (0,0) -- (-0.87,-0.5);
  \node at (-0.55,-0.55) {\scriptsize$\beta, m$};

  \draw[decorate, decoration={snake, amplitude=1mm, segment length=3mm}] (0,0) -- (0.87,-0.5);
  \node at (0.55,-0.55) {\scriptsize$\gamma, n$};

  \node at (0.2,0.24) {\scriptsize$\mu,i$};
  \node at (-0.15,0.24) {\scriptsize$\sigma$};
  \node at (-0.2,-0.3) {\scriptsize$\nu,j$};
  \node at (0.2,-0.3) {\scriptsize$\rho,k$};

  \draw[midarrow, draw=none] (90:1) arc[start angle=90, end angle=-30, radius=1];
  \draw[midarrow, draw=none] (210:1) arc[start angle=210, end angle=90, radius=1];
  \draw[midarrow, draw=none] (330:1) arc[start angle=330, end angle=210, radius=1];

\coordinate (A) at (0,1);            
  \coordinate (B) at (-0.87,-0.5);     
  \coordinate (C) at (0.87,-0.5);      
  \coordinate (O) at (0,0);            

\node[circle, fill=white!20, draw=blue, inner sep =2, align=center] at (O) {\scriptsize{4}};
\node[circle, fill=white!20, draw=blue, inner sep =2, align=center] at (A) {\scriptsize{1}};
\node[circle, fill=white!20, draw=blue, inner sep =2, align=center] at (B) {\scriptsize{2}};
\node[circle, fill=white!20, draw=blue, inner sep =2, align=center] at (C) {\scriptsize{3}};

  \node[circle, fill=white!20, 
  inner sep =0, align=center] at (0.0,.24) {\small $\ast$};

  \node[text=red] at (30:1.12) {\scriptsize 6};
  \node[text=red] at (150:1.12) {\scriptsize 4};
  \node[text=red] at (270:1.12) {\scriptsize 5};
  
  \node[text=red] at (-.15,.5) {\scriptsize 1};
  \node[text=red] at (-.5,-.15) {\scriptsize 2};
  \node[text=red] at (.55,-.15) {\scriptsize 3};

\end{tikzpicture}
\end{center}
\caption{An example of a three loop graph in a QCD-like theory. The asterisk denotes a covariant derivative of a fluctuation field (here with respect to $y_4^\sigma$).}
\label{fig:example}
\end{figure}

In order to show more explicitly the formalism at work, consider the three-loop diagram in figure \ref{fig:example}. 
From its definition eq.~(\ref{eq:Gammadef}) one can directly write down the function $\Gamma$ for this diagram, it reads
\begin{align*}
\Gamma(\t_i,y_n,k_i)
={}&-\tr \biggl\{ \biggr.
\underbrace{\bigl[g\,e(y_1)e\indices{_a^\alpha}(y_1)\gamma^a\mathfrak t^\ell\bigr]}_{C_1}
\underbrace{(\sl k_4+i\sl D_1+m)B(\t_4,X_\frac12,y_1,y_2)}_{E_4}
\\
{}&
{}
\underbrace{\bigl[g\,e(y_2)e\indices{_b^\beta}(y_2)\gamma^b\mathfrak t^m\bigr]}_{C_2}
\underbrace{(\sl k_5+i\sl D_2+m)B(\t_5,X_\frac12,y_2,y_3)}_{E_5}
\\
{}&
\underbrace{\bigl[g\,e(y_3)e\indices{_c^\gamma}(y_3)\gamma^c\mathfrak t^n\bigr]}_{C_3}
\underbrace{(\sl k_6+i\sl D_3+m)B(\t_6,X_\frac12,y_3,y_1)}_{E_6}
\biggl.\biggr\}
\\
{}&
	\underbrace{(D_{4}^\sigma-ik_1^\sigma)[-B(\t_1,X_1,y_4,y_1)]\indices{^{\mu i}_{\alpha\ell}}}_{E_1}
\\
{}&
	\underbrace{[-B(\t_2,X_1,y_4,y_2)]\indices{^{\nu j}_{\beta m}}}_{E_2}
	\underbrace{[-B(\t_3,X_1,y_4,y_3)]\indices{^{\rho k}_{\gamma n}}}_{E_3}
\\
{}&	
\underbrace{g\,e(y_4)f_{ijk}\bigl[g_{\mu\nu}(y_4)g_{\sigma\rho}(y_4)-g_{\mu\rho}(y_4)g_{\sigma\nu}(y_4)\bigr]}_{C_4}
e^{-(\t_4+\t_5+\t_6)m^2}
\end{align*}
Here we defined $e(y)\equiv\det e\indices{_\mu^a}(y)$, $g$ is the gauge coupling,  and the overall minus sign results from the closed Fermion loop. The trace is over the Dirac and gauge group indices in the representation of the Fermions, whose generators we denote by $\mathfrak t^i$. 
Note that the covariant derivatives $\sl D$  contain further factors of the vielbein, as well as gauge and spin connections that have to be expanded as well.
Even though the expansion of the six $B$ functions as well as 
various vielbeins and their determinants will result in rather lengthy intermediate expressions they are in principle completely straightforward.

The other ingredient in our master formula eq.~(\ref{eq:master}) is the function $I'(\t_i,p_n,z_i)$, which we reiterate, is independent of any background fields.
For our tetrahedron topology it is given by a rather complicated expression; instead of reproducing it here we give only the relevant incidence matrix
\be
\B=
\begin{pmatrix}
1&0&0&1&0&-1\\
0&1&0&-1&1&0\\
0&0&1&0&-1&1\\
-1&-1&-1&0&0&0
\end{pmatrix}\,,
\ee
which can be used to calculate $I'(\t_i,p_n,k_i)$ with the code given in appendix \ref{sec:momentumintegral}. Notice that the direction of momentum flow is arbitrary for real fields, while for complex fields it is conveniently taken parallel to particle number flow.

\section{Conclusions}
\label{sec:conclusions}

In this paper we have presented a general method for computing the contribution of arbitrary multi-loop Feynman graphs in general gauge and gravitational backgrounds, for fields of spin $\leq 1$. The main technical result is the expansion of the (normalized) HK $B(\t, X,y,y')$ in RNCs and radial gauge, as summarized in eq.~(\ref{eq:Bfinal}). 
The complexity of the  expressions increases with operator dimension even faster 
than the usual diagonal HK coefficients.
 Therefore, beyond dimension four, some kind of automated
code seems to be the only feasible choice. We leave the implementation of such a code to future work.

Once the expansion of the function $B(\tau,X,y,y')$ is known, any multi-loop diagram is a straightforward application of our master formula eq.~(\ref{eq:master}). This formula depends on the function $I'(\t_i,p_n,z_i)$, which was calculated in ref.~\cite{vonGersdorff:2023lle} in terms of graph polynomials and is summarized
in appendix \ref{sec:momentumintegral}.

Of course, at the end one still has to perform the integral over the Schwinger-Feynman parameters, which is equivalent to the reduction to master-integrals of the vacuum diagrams at the required loop order. These integrals are fully known up to three loop order (see for instance ref.~\cite{Martin:2016bgz}).

\section*{Acknowledgements}
GG acknowledges financial support by the Conselho Nacional de Desenvolvimento
Científico e Tecnológico (CNPq) under fellowship number 313238/2023-5, as well as 
the Funda\char"00E7ão de Amparo à Pesquisa do Estado do Rio de Janeiro (FAPERJ) under 
project number 210.785/2024. 
IC acknowledges a financial support by IP Paris/École Polytechnique.

\appendix

\section{Notation and conventions}

\label{sec:conventions}

Our metric has signature $+---$. Our sign conventions for the curvature amount to  
\begin{align}
\Gamma^\mu_{\r\s}&=\tfrac{1}{2}g^{\m\a}(g_{\r\a,\s}+g_{\s\a,\r}-g_{\r\s,\a})\,,\nn\\\
R^\m_{\ \n\r\s}&=\Gamma^\m_{\n\s,\r}-\Gamma^\m_{\n\r,\s}+\dots
\end{align}
For the Ricci tensor and scalar curvature, we use $R^{}_{\mu\nu}=R^\r_{\ \m\r\n}$, $R=R^{\m}_{\ \m}$.\footnote{In the classification of ref.~\cite{Misner:1973prb} this corresponds to the $(-++)$ sign convention.}
We denote curved indices by lowercase Greek letters and local Lorentz indices by lowercase Latin letters.


We use the following convention for the connections appearing in the covariant derivative
\be
D_\mu=\nabla_\mu+\mathcal A_\mu+\omega_\mu\,,
\ee
where $\nabla$ is covariant with respect to curved indices only and $\mathcal A_\mu$ and $\omega_\mu$ are the matrix-valued gauge and spin connections respectively.
Our conventions for the structure constants read
\be
[\mathfrak t_i,\mathfrak t_j]=f\indices{_{ij}^k} {\mathfrak t_k},
\ee
such that the adjoint generators are $(\mathfrak t^{\rm adj}_i)\indices{^k_j}=f\indices{_{ij}^k}$.


%

\section{The universal momentum integral}
\label{sec:momentumintegral}

In reference \cite{vonGersdorff:2023lle}, we computed the Fourier transform of the function 
\be
I(\t_i,y_n,k_i)\equiv (-i)^L \delta(\bar y)e^{-iy\tra\B k+ k\tra\T_0k}\,,
\ee
where $\bar y\equiv \frac{1}{V}\sum_{n=1}^V y_n$, $\B$ is the directed incidence matrix of the graph
\be
\B_{ni}=\begin{cases}
+1& $if edge $ i $ enters vertex $ n \\
-1& $if edge $ i $ leaves vertex $ n \\
0 & $else$
\end{cases}
\label{eq:Bni}
\ee
and $\T_0=\diag(\t_i)$. Then, 
\be
I(\t_i,p_n,z_i)=(4\pi)^{-\frac{dL}{2}}\Delta^{-\frac{d}{2}}\exp(\tfrac{1}{4}z\tra \Q z-iz\tra \R p-p\tra \U p)\,,
\ee
where $\Delta$ is the first Symanzik polynomial of the graph and $ \Q$, $ \R$ and $\U$ are 
matrices whose components are polynomials in the $\t_i$, divided by $\Delta$.
We refer the reader to reference \cite{vonGersdorff:2023lle} for the definition and characterization of these
matrices. 
In the following we give some simple Mathematica code for their calculation.\footnote{In the code, $\B$ is the incidence matrix of the two-loop sunset vacuum graph, it can be replaced with any connected vacuum graph.}  
\begin{verbatim}
B = {{1, 1, 1}, {-1, -1, -1}};
{NV, NP} = Dimensions[B];
T0 = Table[Subscript[t, i], {i, 1, NP}] // DiagonalMatrix;
T = T0 + Transpose[B] . B/x;
Delta = SeriesCoefficient[Det[T]/NV, {x, 0, 1 - NV}];
Q = SeriesCoefficient[Inverse[T], {x, 0, 0}];
R = SeriesCoefficient[Inverse[T] . Transpose[B], {x, 0, 1}];
U = -Transpose[R] . T0 . R;
\end{verbatim}
The code assumes that the diagram is connected, and that columns of $\B$ corresponding to 
self-loops (edges that start and end at the same vertex) are identically zero.

Finally the quantity $I'$ can be obtained from $I$ by replacing $p_1$ by $-p_2-\dots-p_V$.

\section{The purely gravitational heat kernel}
\label{sec:grav}

In this appendix we reproduce for completeness the expansion of $B(\t,X,y,y')$ around $\t=0$, $y=0$, and $y'=0$ for the case of vanishing spin and gauge connections, as performed in \cite{Carneiro:2024slt}. We organize the result by dimensions of the operators, keeping up to dimension four,
\be
B^\text{grav}=
B^\text{grav}_0
+B^\text{grav}_1
+B^\text{grav}_2
+B^\text{grav}_3
+B^\text{grav}_4\,,
\ee
where  
\begin{align}
 B^\text{grav}_0={}&1\,,
\\
 B^\text{grav}_1={}&0\,,
\\
 B^\text{grav}_2={}&
	- \tfrac{\t}{6}R-\t X
	+ \tfrac{1}{12}R_{\mu\nu} 
		\left(y^\m y^\n+y'^\m y'^\n\right)
	-\tfrac{1}{6}R_{\mu\nu} \,
		y^\m y'^\n 
	-\tfrac{1}{12\t}R_{\m\r\n\s}\, 
		y^\m y^\n y'^\r y'^\s \,,
\\
 B^\text{grav}_3={}&	
	 \left(
	-\tfrac{\t}{12}R_{;\mu}
	-\tfrac{\t}{2}X_{;\mu}\right)
		\left(y^\m+y'^\n\right)
	+\tfrac{1}{24}R_{\mu\nu;\rho}\,
		\left(y^\m y^\n y^\r +y'^\m y'^\n y'^\r\right)
	\nn\\
	&+\left(\tfrac{1}{12}R_{\mu\nu;\rho}-\tfrac{1}{6}R_{\rho\mu;\n}\right)
		\tfrac{1}{2}\left(y^\m y^\n y'^\r+y'^\m y'^\n y^\r\right)
	-\tfrac{1}{24\t}R_{\m\r \n\s;\t}\,
		y^\mu y^\nu(y^\t+y'^\t) y'^\r y'^\s \,,	
\\
%
 B^\text{grav}_4 ={}&	
	\ \tfrac{\t^2}{30}\nabla^2 R
	+\tfrac{\t^2}{72}R^2
	-\tfrac{\t^2}{180}R_{\m\n}R^{\m\n}
	+\tfrac{\t^2}{180}R_{\m\n\r\s}R^{\m\n\r\s}
	+\tfrac{\t^2}{6}\nabla^2 X
	+\tfrac{\t^2}{6}RX
	+\tfrac{\t^2}{2}X^2
	\nn\\&
	+\left(	
	-\tfrac{\t}{20}R_{;\mu\nu}
	-\tfrac{\t}{60}\nabla^2R_{\mu\nu}
	-\tfrac{\t}{3}X_{;\mu\nu}
	-\tfrac{\t}{6}XR_{\mu\nu}
	+\tfrac{\t}{45}R^\a_{\ \mu}R^{}_{\a\nu}
	\right.
\nn\\&
	\phantom{+}
	\left.
	-\tfrac{\t}{36}RR_{\mu\nu}
	-\tfrac{\t}{90}R^{\a\g}R_{\a\mu\g\nu}
	-\tfrac{\t}{90}R^{\a\g\kappa}_{\ \ \ \ \mu}	R^{}_{\a\g\kappa\nu}
	\right)
		\tfrac{1}{2}\left(y^\mu y^\nu+y'^\m y'^\n\right)
	\nn\\
	&+\left(
	-\tfrac{\t}{30}R_{;\mu\nu} 
	+\tfrac{\t}{60}\nabla^2R_{\mu\nu}
	-\tfrac{\t}{6}X_{;\mu\nu}
	+\tfrac{\t}{6}XR_{\mu\nu}
	-\tfrac{\t}{45}R^\a_{\ \mu}R^{}_{\a\nu}
	\right.
	\nn\\&
	\phantom+
	\left.
	+\tfrac{\t}{36}RR_{\mu\nu}
	+\tfrac{\t}{90}R^{\a\g}R_{\a\mu \g\nu}
	+\tfrac{\t}{90}R^{\a\g\kappa}_{\ \ \ \ \mu}R^{}_{\a\g\kappa\nu}
	\right)
		y^\mu y'^\nu
	\nn\\	
	&+
	\left(	
	\tfrac{3}{10}R_{\mu\nu;\rho\sigma}
	+\tfrac{1}{15}R_{\a\mu\nu}^{\ \ \ \ \g}R^\a_{\ \rho\sigma\g}
	+\tfrac{1}{12}R_{\mu\nu}R_{\rho\sigma}\right)
		\tfrac{1}{24}\left(y^\m y^\n y^\r y^\s+y'^\m y'^\n y'^\r y'^\s\right)
	\nn\\
	&+\left(
	\tfrac{1}{120}R_{\mu\nu;\rho\sigma}
	-\tfrac{3}{20}R_{\s\m;\n\r}
	+\tfrac{11}{120}R_{\m\n;\s\r}
	\right.
	\nn\\
	&
	\phantom{+}
	\left.
	-\tfrac{1}{15}R_{\a\mu\nu}^{\ \ \ \  \g}R^\a_{\ \r\s \g}
	-\tfrac{1}{12}R_{\mu\nu}R_{\rho\sigma}
	\right)
		\tfrac{1}{6}\left(y^\m y^\n y^\r y'^\s+y'^\m y'^\n y'^\r y^\s\right)
	\nn\\
	&+
	\left(
	\tfrac{19}{180}R_{\m\n;\r\s}
	+\tfrac{19}{180}R_{\r\s;\m\n}	
	-\tfrac{11}{90}R_{\m \r;\n \s}
	-\tfrac{11}{90}R_{\r\m;\s \n}
 	- \tfrac{1}{18} \nabla^2 R_{\m\r\n\s}
	\right.\nn\\
	&\left.
	\phantom{+}
	-\tfrac{11}{90}R_{\m\r\a\g}R_{\n\s}^{\  \ \,\a\g}
	+\tfrac{7}{45}
	R_{\a\m\r}^{\ \ \ \  \g}R^\a_{\ \n\s \g}
	-\tfrac{4}{45}
	R_{\a\mu\nu}^{\ \ \ \  \g}R^\a_{\ \r\s \g}
	\right.\nn\\
	&\left.
	\phantom{+}
	+\tfrac{1}{36}R_{\m\n}R_{\r\s}
	+\tfrac{1}{18}R_{\m\r}R_{\n\s}
	\right)
		\tfrac{1}{4}y^\m y^\n y'^\r y'^\s
	\nn\\
	{}&
		-
	\left(
	\tfrac{1}{3\t}R_{\m\r\n\s}R_{\k\t}
	\right.	
	\left.
	+\tfrac{4}{15\t}R^{\alpha}_{\ \k\m\r}R_{\alpha\t\n\s}^{}
	+\tfrac{1}{5\t}R_{\mu\r\nu\s;\k\t}\right) 
		\tfrac{1}{48}\, y^\m y^\n (y^\k y^\t+y'^\k y'^\t)y'^\r y'^\s
	\nn\\&	
	+\left(
	\tfrac{1}{2\t}R_{\m\n\r\s}R_{\k\t}
	-\tfrac{4}{5\t}R^{\alpha}_{\ \mu\nu\t}R_{\alpha\r\s\k}^{}
	\right.	
	\left.
	-\tfrac{3}{40\t}R_{\mu\r\nu\s;\k\t}
	-\tfrac{3}{40\t}R_{\mu\r\nu\s;\t\k}
	\right)
		\tfrac{1}{36}\,y^\mu y^\nu  y^\k y'^\r y'^\s y'^\t 
	\nn\\&
	+\tfrac{1}{144\t^2}R_{\m\r\n\s}R_{\t\k\l\eta}\,
		y^\m y^\n y^\t y^\l y'^\r y'^\s y'^\k y'^\eta\,.	
\label{eq:B4}
\end{align}

Notice that when applied to eq.~(\ref{eq:Bfinal}), $X$  transforms under gauge (including local Lorentz) transformations, and hence the covariant derivatives on $X$ become covariant with respect to those transformations also.

\bibliographystyle{JHEP}

\bibliography{literature}

\end{document}